\documentclass[fleqn,10pt]{wlscirep}
\usepackage[utf8]{inputenc}
\usepackage[T1]{fontenc}

\title{Occupant Privacy Perception, Awareness, and Preferences in Smart Office Environments}

\author[1]{Beatrice Li}
\author[2]{Arash Tavakoli}
\author[1,*]{Arsalan Heydarian}
\affil[1]{Department of Engineering Systems and Environment, University of Virginia, Charlottesville, VA 22904, USA}
\affil[2]{Department of Civil and Environmental Engineering, Stanford University, Stanford, CA 94305, USA}

\affil[*]{ah6rx@virginia.edu}


\begin{abstract}
Building management systems tout numerous benefits, such as energy efficiency and occupant comfort but rely on vast amounts of data from various sensors. Advancements in machine learning algorithms make it possible to extract personal information about occupants and their activities beyond the intended design of a non-intrusive sensor. However, occupants are not informed of data collection and possess different privacy preferences and thresholds for privacy loss. While privacy perceptions and preferences are most understood in smart homes, limited studies have evaluated these factors in smart office buildings, where there are more users and different privacy risks. To better understand occupants' perceptions and privacy preferences, we conducted twenty-four semi-structured interviews between April 2022 and May 2022 on occupants of a smart office building. We found that data modality features and personal features contribute to people's privacy preferences. The features of the collected modality define data modality features -- \textit{spatial, security, and temporal context}. In contrast, personal features consist of one's awareness of data modality features and data inferences, definitions of privacy and security, and the available rewards and utility. Our proposed model of people's privacy preferences in smart office buildings helps design more effective measures to improve people's privacy.
\end{abstract}

\begin{document}

\flushbottom
\maketitle

\thispagestyle{empty}

\section*{Introduction}\label{intro}
With the increasing number of Internet of Things (IoT) devices in smart office buildings, building management systems (BMS) are utilized to improve energy efficiency and occupant comfort by automatically managing indoor environmental conditions, such as temperature, humidity, and lighting conditions, based on occupant behavior and preferences \cite{Saputro2016PrivacyBuildings, Labeodan2015OccupancyEvaluation, Molina-Solana2017DataReview, Awada2021HomeIEQWorker, Aryal2018SmartDesk, Aryal2018EnergyComfortOffice, Simin2018ActivityEnergyBuild, Ghahramani2018UnsupervisedThermal}. As research in this area has been increasing over recent years, there is a clearer need to understand occupant behavior and human-building interaction \cite{Becerik-Gerber2022TenQuestionsHBI,Becerik-Gerber2022HBI}. While increased research on occupant behavior monitoring has led to significant insights into building energy management, occupant comfort, and well-being, most, if not all, of these objectives are founded upon collecting data from the occupants within the built environment. 

The sensors embedded in smart buildings (including commercial and residential buildings) fall into environmental sensing, occupancy detection and energy sensing categories with varying levels of intrusiveness \cite{Abade2018AEnvironments, Weng2012FromEfficiency, Lee2019ExploringBuildings}. However, despite the categories of sensors and whether they are ``intrusive'' or ``non-intrusive,'' through advancements in machine learning and signal processing, personal information can be inferred from data collected by the so-called non-intrusive sensors \cite{Kroger2019UnexpectedThings}. A study found that most participants believed in some monitoring and tracking at work, while some explicitly expressed concern about data inferences that would reveal personal information \cite{Harper2020UserBuildings}. Occupants may not even be aware of what information is being collected and how it is used as consent is often implicit \cite{Apthorpe2018DiscoveringIntegrity, Pathmabandu2020AnBuildings}. For instance, an indoor air quality sensor, designed to monitor carbon dioxide ($CO_2$) and total volatile organic compounds (TVOC) levels in the space for the health of occupants, can reveal information such as whether someone is in the space, how many people are present, what general activities they might be doing (e.g., in a meeting, eating, etc.) and when they arrive and leave their space \cite{Wu2021TheChallenge, Pathmabandu2020AnBuildings, Gorjani2022SmartHomeActivity, Kessler2020PrivacyBuildings}. Such inferences of occupant activities can exponentially grow as more modalities are fused together. The lack of awareness of data inferences coupled with implicit consent means that people using a smart office building space lack control of their privacy. In fact, a study found that most participants believed in some monitoring and tracking at work, while some explicitly expressed concern about data inferences that would reveal personal information \cite{Harper2020UserBuildings}. The concern of data inferences is contrasted by interviews conducted by Zheng et al. \cite{Zheng2018UserPrivacy}, which found skepticism for privacy risks posed by non-audio or video (A/V) data in smart homes.

The lack of awareness extends to the data collection processes as well. A previous study by Harper et al. found that while people were aware of some data collection, they lacked confidence in their knowledge of the process in smart buildings \cite{Harper2020UserBuildings}. Another study suggested that there are different areas of privacy sensitivity in smart homes compared to smart buildings due to \textit{potential embarrassment} in homes and the addition of \textit{potential consequences} in buildings \cite{McCreary2016TheBuildings}. \textit{Potential consequences} can be explained through the concern that people in positions of authority can use the data against their subordinates \cite{Harper2022CommercialSB, Lee2018SensingUtility}. The different user behaviors and power dynamics between occupants of a smart office building motivates further research into user perceptions of privacy within these buildings. 

With the complex nature of privacy, many studies are conducted to define privacy norms despite no central definition of privacy \cite{Apthorpe2018DiscoveringIntegrity, McCreary2016TheBuildings}. Nissenbaum developed the theory of privacy as contextual integrity, where the protection of privacy is tied to social contexts and the norms defined within it \cite{Nissenbaum2004LawReview, Nissenbaum2009PrivacyLife}. It posits that to evaluate the preservation of privacy, there are five parameters of information flows to consider -- \textit{the information subject, information type, sender, recipient, and transmission principle}. It has been shown that context plays an important role in user perceptions and attitudes toward privacy \cite{McCreary2016TheBuildings}. Thus, it is important to elucidate the components that contribute to the privacy sensitivity of an area and the contexts in smart office buildings. 

Previous works were primarily in smart homes, and those in smart office buildings were mostly through surveys that did not allow for the questioning of participants. Additionally, for studies on smart homes, the findings may not translate to smart office buildings wholly. Our study builds on the previous works by conducting semi-structured interviews that thoroughly assess the perceptions of privacy and the mental models of data collection and usage in a smart office building space. This study focuses on enhancing the knowledge of the aforementioned gaps by performing a detailed interview with the occupants of a smart office building. 

Consequently, this study aims to answer the following research questions:
\begin{itemize}
    \item \textbf{RQ1:} What contexts are people sensitive in sharing their data?
    \item \textbf{RQ2:} How does perceived potential benefit/reward alter privacy attitudes?
    \item \textbf{RQ3:} How do perceptions of smart office building IoT privacy align with reality?
\end{itemize}

\section*{Results}\label{results}
This section describes the components influencing occupants' privacy preferences in a smart office building from semi-structured interviews with 24 participants. Two components contributing to data privacy preferences emerged from our interviews: \textit{data modality features} and \textit{personal features}. Data modality features focus on the context surrounding data collection with the respective modality. In contrast, personal features are the individual's values and the mental models of data collection that influence their privacy preferences. 

\begin{figure}[ht]
  \centering
  \includegraphics[width=0.4\linewidth]{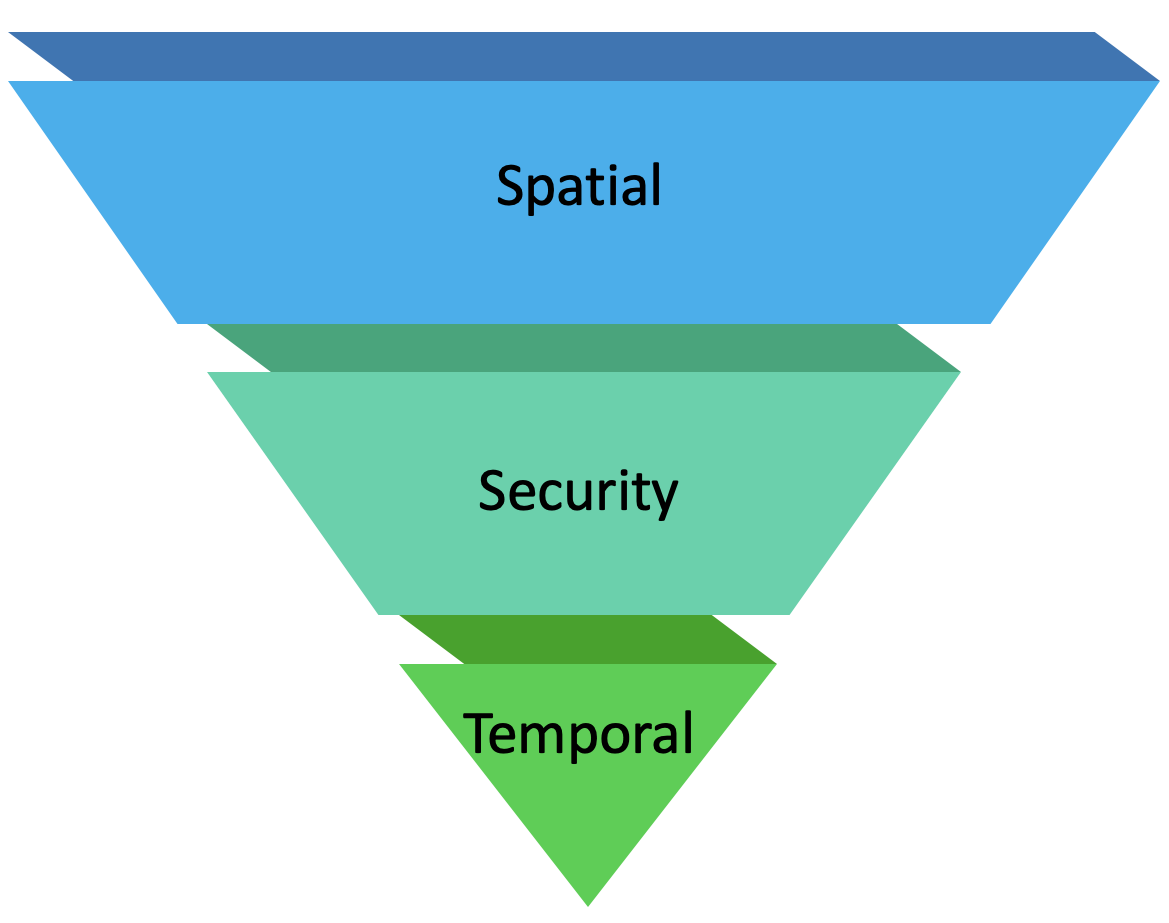}
  \caption{There is a hierarchy of data modality features that influence people's privacy preferences in smart buildings. Spatial context is the most influential, with decreasing influence from top to bottom.}
  \label{fig:context-pyramid}
\end{figure}

\subsection*{Data Modality Features}
People's willingness for data collection depended largely on 1) the modality being collected and 2) the features of that modality, which include \textit{spatial, security, and temporal context}. Modality is the different types of data collected (i.e., environmental, audio, and video). For each modality, spatial context is defined as the physical space and who is in that space. In the setting of this study, an individual's workspace could be a desk with dividers in an open-plan office or an office with transparent glass walls designed for a single occupant. Security context refers to the data access protocols, encryption protocols, and the level of anonymity. There are two aspects to the temporal context: the time of day and the age of the data being used. The interviews have shown a hierarchical order to the data modality features where participants have a specific privacy preference given each modality's \textit{spatial, security, and temporal context}, as seen in Figure \ref{fig:context-pyramid}.

\subsubsection*{Modality Sensitivity Scale}
\begin{figure}[ht]
  \centering
  \includegraphics[width=0.5\linewidth]{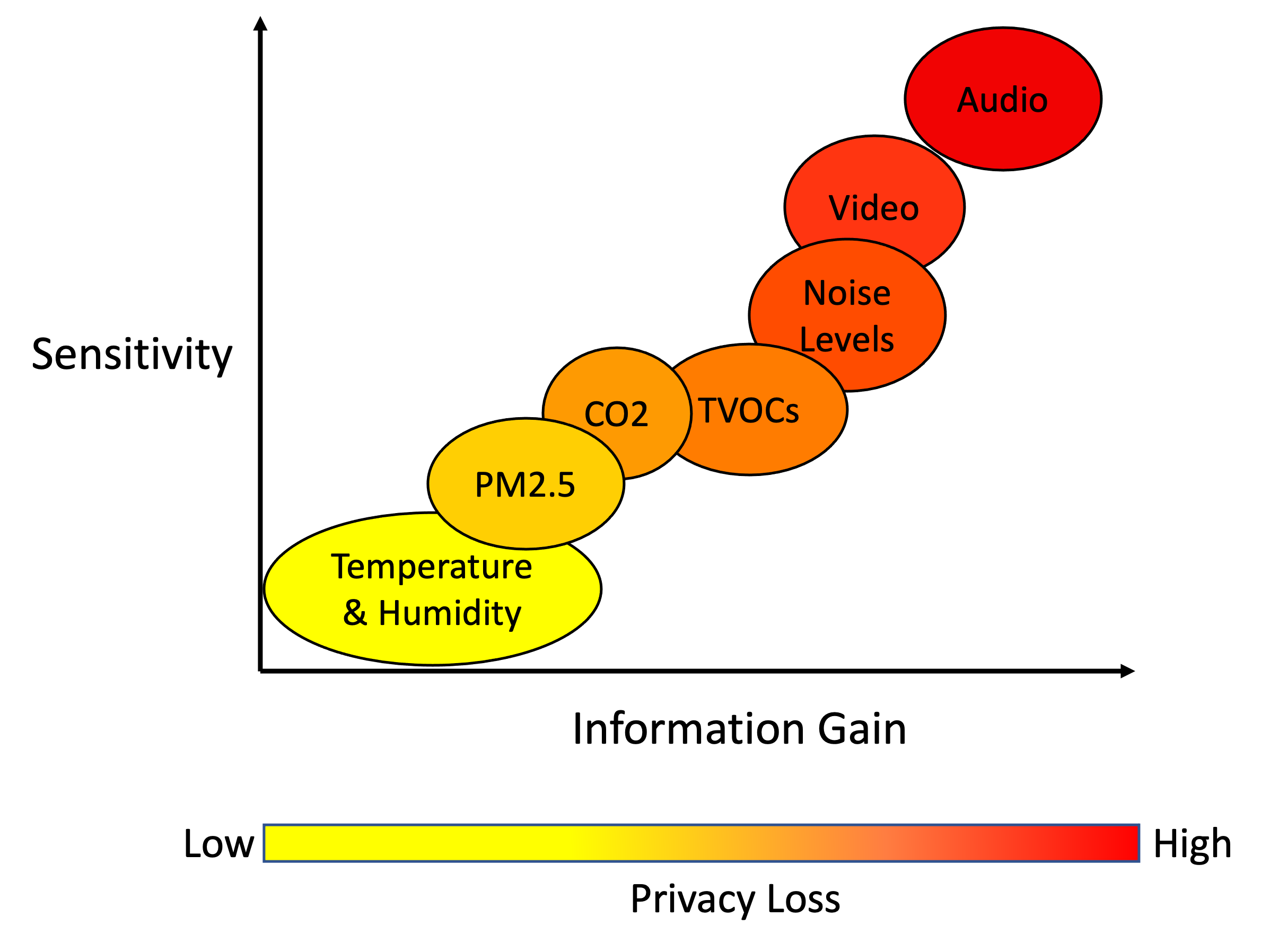}
  \caption{Participant responses have shown different levels of sensitivity towards the various data modalities. The data modalities can be plotted on a scale comparing sensitivity to the information richness of the data.}
  \label{fig:sensitivity}
\end{figure}

For the purposes of the study and based on the responses, modality can be grouped into environmental, audio, and video data categories. There is a sensitivity scale for the data modalities where our participants are less sensitive and more comfortable with environmental data being collected, shown in Figure \ref{fig:sensitivity}. In contrast, people are most sensitive to audio recordings, while video data also has high sensitivity but not as much as audio recordings. Noise levels are a part of audio data but are viewed as less intrusive as compared to recordings of conversations. Participants indicated they were more comfortable and accepting of noise levels rather than audio recording data:

\begin{quote}
    \textit{"I feel like noise level would be okay. But I feel like, uh, audio recording, like in confined spaces, maybe not."} (P9)
\end{quote}

\noindent When asked about what data modality they would be okay with and where many cited A/V recording data as being intrusive and feelings of discomfort:

\begin{quote}
    \textit{"But I wouldn't want like recordings of audio from really anywhere because you have conversations with people. And if you feel like all your conversations are being recorded, that would be kind of creepy."} (P8)
\end{quote}

\noindent A/V data appeared to be the bottom line for many people and had the most restrictions compared to environmental data, particularly audio recordings. Individual sentiments towards data collection that borders on surveillance are negatively viewed. However, there is also a spatial context component in the sensitivity of the data modality toward deeming what data collection is acceptable.

\subsubsection*{Spatial Context}
Spatial context is defined as where the individual is, who they interact with at the time of data collection, and what activity they are engaged with. Many participants define their data collection preferences with respect to the spatial context -- primarily by their designated workspace or public areas -- and the modality sensitivity scale. 

Half (\textit{n}=12) of the participants said they would not be okay with A/V recording at their workspace. In comparison, only one individual indicated that they are okay with audio recording at their workspace because they do not talk much there. Even in open-plan offices, individual workspaces are valued and considered personal spaces. Two participants that stipulated they would not be okay with A/V at their workspace also preferred not to have any A/V data collected in individually occupied spaces around the building (e.g., single occupancy offices), including common spaces, if they were the only person there. This posits that public common areas with a single occupant are considered more private overall. Nearly everyone (\textit{n}=23) was okay with environmental data collected everywhere except for one individual, who was not okay with any data collected at the workspace, which conveys the acceptable nature of environmental data.

Approximately 54\% (\textit{n}=13) answered that they would be okay with either audio or video in other more public spatial contexts, including corridors, common spaces, and conference rooms. Three individuals that were okay with video data being collected cited safety and protection of valuables as the primary purpose:

\begin{quote}
    \textit{"I can understand that like video is kind of helping with like, surveillance is kind of safety."} (P22)
\end{quote}

\noindent Others described that audio or video, or both, were okay in more public spatial contexts because it is in the presence of others:

\begin{quote}
    \textit{"I feel fine about [audio and video] being collected in the [public] ... I'm normally aware that I'm in a more public area, whereas if I'm in my actual office, I don't know how I feel about data being collected, audio data and video data, in my office."} (P15)
\end{quote}

\noindent The underlying justification is that when one speaks in a common space, others can also hear it, which can be equated to "something or someone" listening in through audio recordings. People are cognizant when in a public space and recognize that they adopt a filter for what they say. 

\subsubsection*{Security Context}
Security context can be defined as who has access to the data and the level of anonymity in the data. The level of anonymity refers to how aggregated the data is or the ease of de-identification. In expressing preferences for data collection, some said they are only okay with recording audio data if it can be aggregated into sound levels, which goes back to the modality sensitivity scale. Over 62\% (\textit{n}=15) explicitly stated that they did not want any identifiable data to be collected. Over 29\% (\textit{n}=7) of participants expressed the preference of being able to control who has access to data, while the majority (\textit{n}=20) were okay with researchers and relevant parties approved by the IRB to have access to their data. Even among those that did not explicitly express approval of researcher use, they indirectly approved as they were okay with residents and people relevant to the smart building (i.e., facility management, researchers) having access to the data. 

Aside from who has access and the level of anonymity, another aspect of security is associated with data storage protocols and how access is managed. Just over half (\textit{n}=13) of the interviewees discussed data security, but only a few (\textit{n}=3) expressed security measures as a prerequisite or as part of their consideration in data collection. The interviews showed that participants cared most about who has access and the level of anonymity, while data storage and how data can be accessed are not thought of as much. 

\subsubsection*{Temporal Context}
Temporal context is defined as when the data collection occurs and the age of data. 75\% (\textit{n}=18) of participants did not think that the age of data matters, but some indicated that using older data is more acceptable. The age of data did not change how acceptable it was for A/V data to be collected, as a few expressed fear of harmful consequences. However, some participants expressed that A/V data may be more acceptable outside work hours. One individual justified video being acceptable outside of work hours to monitor the equipment within the space in case of intruders. 

\subsection*{Personal Features}
Other than data modality features, personal features influence individuals' privacy preferences and acceptability towards data collection. Specifically, the interviews showed definitions of privacy and security, awareness, and rewards as the primary features that vary among individuals regarding privacy preferences.

\subsubsection*{Participants' Definition of Privacy and Security}
The participants often blurred the distinctions between privacy and security in the smart office spaces, and most did not have readily available definitions. However, providing examples as part of their responses helped verbalize their definitions, which goes on to demonstrate that inability to define privacy does not equate to not caring about privacy. Most defined privacy as data that is not identifiable:

\begin{quote}
    \textit{"As long as the information cannot be used to identify specific person..."} (P3)
\end{quote}

\noindent One participant also recognized that the definition of privacy is subject to change due to the technologies and their capabilities:

\begin{quote}
\textit{"This is an ever evolving space, so we have to deal with issues as they come up right now. I think that definition is a pretty issue free one, but we will see in the future so that's just why I say for now."} (P4)
\end{quote}

\noindent The definitions of privacy varied, but almost all revolved around information tied to the individual or space belonging to the individual. When asked for their definition of security in a smart office space like the one they work in, some participants wanted clarification on whether it was with respect to data security or building security. In response, we left that up to the participants to determine what they think is most suitable. Some participants only provided a definition that references the physical security and safety of the building, while data was not discussed:

\begin{quote}
    \textit{"That's a hard one for me because I'm in an close space, [so] security is not something that I really worry about...I don't know how it will change if it's in a smart building or not for me."} (P18)
\end{quote}

\noindent The latter response suggests no difference in the definition of security between a smart building as compared to a conventional building. In contrast, security definitions that fall under data security were primarily about authorized access, preventing unauthorized users from accessing data, and encryption. Some participants also expressed that security in the smart office building is composed of two parts -- building security for the safety of occupants and data security for protecting sensitive information:

\begin{quote}
    \textit{"Safety is the word that comes up first [for] security... So security, I think of [is] you know, physical safety and then security [is] also protection against viruses and people hacking the system and things like that."} (P14)
\end{quote}

\noindent As participants were asked for the definitions of privacy and security successively, some participants provided the same response to both questions or noted a minimal difference between the two. Privacy was more focused on the individual and the space they occupy, while security had more to do with the safety of the building. The differences in how participants define privacy and security are also presented in differences in data preferences. 

\subsubsection*{Awareness of Data Modality Features and Inferences}
All participants were aware of environmental data being collected -- namely, $CO_2$, temperature, and humidity. A few indoor environmental quality (IEQ) metrics collected in the space that a few participants did not mention were TVOCs, $PM_{2.5}$, voltage, and noise levels. It should be noted that the awareness of IEQ data collection can be attributed to the display of numerous IEQ sensors around the space if recognized. Beyond these metrics, participants are uncertain if any other data is collected, like A/V data. Noise levels and audio recordings were a point of contention as some participants pointed out that noise levels could mean that audio recordings are collected.

As most participants were concerned with collecting A/V data, almost everyone found environmental data to be completely acceptable. The majority were averse to identifiable data, but it was mostly referring to A/V data as only a few (\textit{n}=3) said they were okay with data inferences from IEQ sensors while even fewer (\textit{n}=4) said they would not want data inferences or at least not without consent or authorization. Most participants were unaware of the risks posed by innocuous sensors, such as IEQ sensors. The contrast is apparent in responses as people oppose A/V data due to tracking behavior but are very accepting of environmental data. One participant defined identifiability as data tagged with the individual's HIPAA identifiers (i.e., name, date of birth, etc.).

The participants that were accepting of environmental data were asked if their response would change if inferences could be made from environmental data, such as their behaviors and routines; there were mixed responses, most with uncertainty. What participants see as identifiable data also seems to vary, but A/V data are indisputably perceived as identifiable. In contrast, one participant supports the use of data inferences to maximize the potential of a smart office building. The privacy risk of data inferences is not as tangible as A/V data, where one would feel tracked and monitored. 

\subsubsection*{Rewards \& Utility}
Through the interviews, numerous responses showed that willingness for data collection depended on the trade-offs. Specifically, participants described a cost-benefit analysis of utility and loss of privacy. About 63\% (\textit{n}=15) of participants indicated that well-being insights have minimal to no impact on their willingness for data collection or that it depended on offered utility. To elaborate, well-being insights can be as simple as providing occupants with information regarding the environment or suggestions on which spaces within the building would be optimal with respect to health, comfort, and productivity. 

The participants also provided a similar answer with respect to the influence of monetary compensation on willingness for data collection:

\begin{quote}
    \textit{"I don't know if financial compensation really quite does it for me because like money is limited. I can't imagine it's gonna be significant."} (P1)
\end{quote}

\noindent However, 33\% indicated that well-being insights would increase their willingness for data collection, including video data, with minimal to no change to the willingness from monetary compensation. However, well-being insights must have a large societal impact, such as cancer diagnosis. Regardless, participants would still want to weigh the costs and benefits. Even as receptiveness to the collection of video data increased with well-being insights offered, some specified that they would want to know how the data is being used and de-identified. One participant's response was the following:

\begin{quote}
    \textit{"If you can detect my audio to tell me that I have cancer, for instance, I'll be like 'yea sign me up for that,' because I think it's really, it does help me, so, no issues and same for video again ... Can you tell me that I have skin cancer just by looking at my face for instance, or can you tell me that I have poor posture looking at my posture and then doing some gait matching. I would be more inclined, but it would not be a yes until I know and I have done my own sort of pros and cons."} (P18)
\end{quote}

\noindent However, one participant actively dedicates time to review privacy policies and settings of every application they use but is willing to provide data when monetary compensation is offered. As evidence, this participant also described his participation in a study that collected extensive physiological data through wearables and frequent surveys. Overall, some reward does incentivize data collection, including individuals who claim to be very private.

Outside of privacy preferences, participants also highly value convenience, and it appeared in many of the interviews as part of their decision-making process -- from purchasing criteria of personal devices (e.g., smartphone, smart home assistants, etc.) to time spent on privacy settings. Even simple strategies and actions that may preserve privacy will be ignored if it becomes a hindrance. For example, one participant used to tape her webcam due to heightened concern about remote spying through cameras but stopped after it proved too bothersome when she needed to use it. As simple as the action was, it conveys the impedance of the smallest inconvenience in implementing privacy protection behaviors. This notion is reinforced when participants are asked how much time and effort they are willing to spend on steps to secure their privacy:

\begin{quote}
    \textit{"I'm sure it is bad, but at least for me, I feel like my desire for privacy is all for convenience. So like the less convenient it is, the more I kind of give on privacy."}
\end{quote}

\noindent Over half of the participants (\textit{n}=15) responded that they would want to spend as little time as possible, and the exact time ranged from 0 minutes to 15 minutes per day at most. They also expressed that they prefer a one-time adjustment rather than daily or frequent adjustments. A few of those participants (\textit{n}=3) went on to elaborate there is no need as there is no threat. In contrast, two participants said they were willing to spend up to 30 minutes, and some (\textit{n}=4) specified 1-2 hours per day. Most participants would rather spend as little time as possible and prefer a one-time adjustment rather than daily or frequent adjustments.

\begin{figure}[ht]
  \centering
  \includegraphics[width=0.75\linewidth]{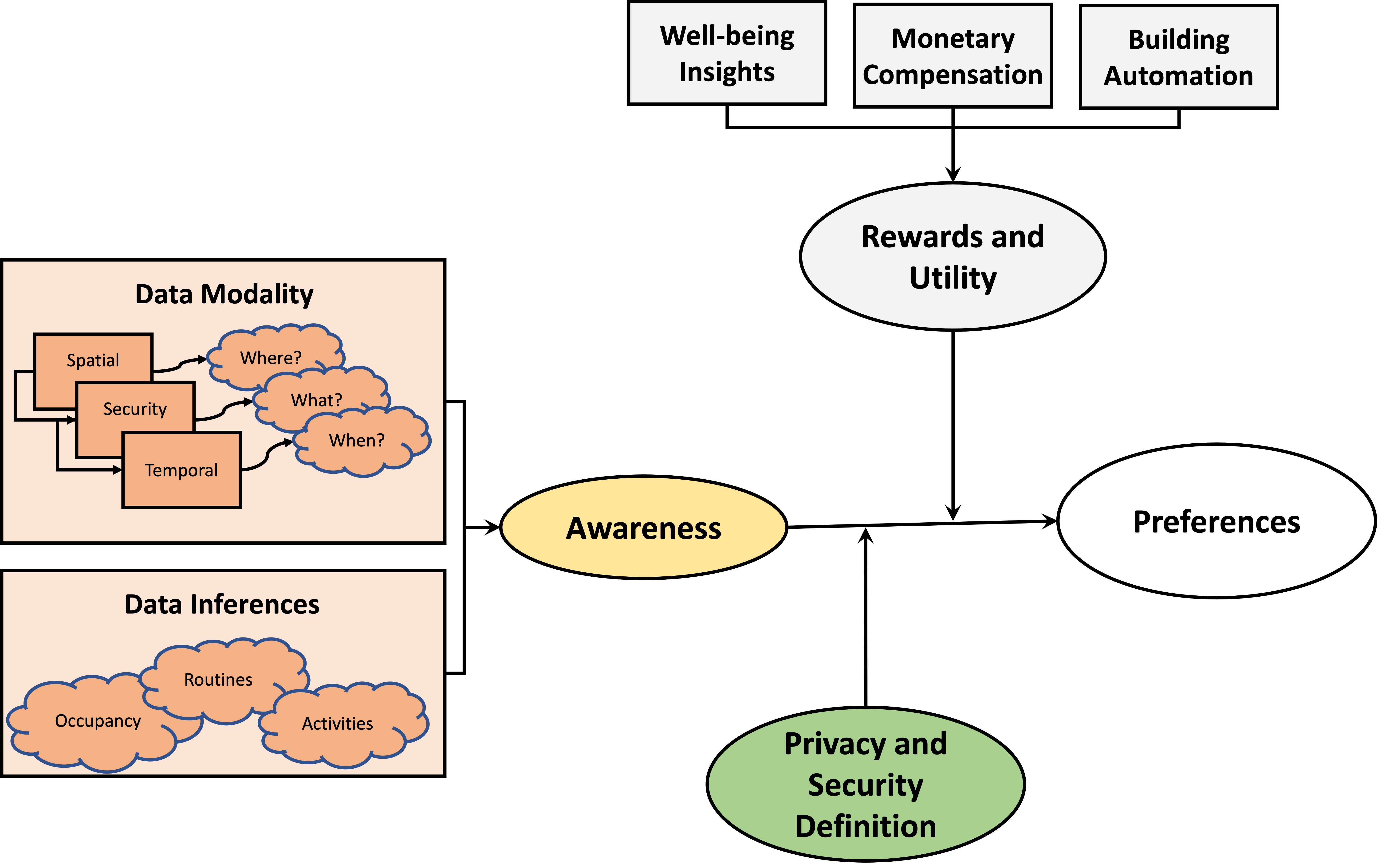}
  \caption{People's privacy preferences are composed of data modality features and personal features. Data modality features and data inferences contribute to the individual's awareness. One's awareness, definitions of privacy and security, and the rewards and utility make up the personal features.}
  \label{fig:preferences}
\end{figure}

\section*{Discussion}\label{discussion}
In our study, we conducted semi-structured interviews among occupants of a smart office building to understand privacy perceptions, awareness, and preferences in a smart office building. Through the analysis of the interviews, we found a gap between the perceptions and reality of smart building IoT privacy, as many were uncertain of the data collection processes and the capabilities of data inferences. The interviews allowed us to understand the underlying mechanisms of privacy preferences, -- personal features, and data modality features. Personal features are intrinsic to the participants: awareness of data modality features and inferences, definitions of privacy and security, and rewards and utility, as shown in Figure \ref{fig:preferences}.

In data modality features, each data modality has its own \textit{spatial, security, and temporal context}, respectively ranked from high to low influence. Spatial context, as the most influential, is evident in how people see their workspace as more private and are less comfortable with intrusive data (i.e., A/V) to be collected compared to more public areas of the building. Our results extend previous findings that people are most sensitive with their personal offices \cite{McCreary2016TheBuildings} to open-plan workspaces. The security context is how easily the data could be de-identified (i.e., noise levels compared to audio) and who has access. The importance of data access can be the concern of \textit{potential consequences} associated with collected data \cite{McCreary2016TheBuildings}. The temporal context is when the data is collected and how old it is. Some responses expressed increased acceptability of A/V data  outside of work hours, which could be due to a lack of people in the space. 

Awareness also extends to data inferences -- the possibilities of inferred information from data. Very few participants brought up data inferences and the respective privacy risks. Even as we prompted the participants with inferences that can be made from the environmental data, such as their routines, they remained either skeptical or unconcerned. The lack of concern stands in stark contrast to a study that demonstrated an accuracy of 74.78\% in predicting activities in smart buildings \cite{Marcello2019Recognition} and 97.8\% in smart homes using innocuous motion and IEQ sensors. The opposition to A/V data is because of its ability to track and monitor people, but similar tracking capabilities can also be found in non-A/V sensors used for building functions. Though A/V data can identify occupants, it is important to recognize that environmental sensors can reveal behaviors and routines, which can be used to track, identify and potentially punish employees. While almost all participants had some knowledge of what data was being collected, they all expressed uncertainty and how it would be used. It becomes apparent that there is a gap between the mental model or awareness and the reality of data collection and privacy risks within the building space, which is in line with previous works \cite{Harper2020UserBuildings,Harper2022CommercialSB}. 

Participants struggle with defining \textit{privacy} and \textit{security} as many blur the two concepts but are able to express their privacy preferences. The interviews frame privacy as context-specific, which aligns with Nissenbaum's theory of privacy as contextual integrity \cite{Nissenbaum2009PrivacyLife}. The differences between smart office buildings and conventional buildings, as participants know them traditionally, are not obvious, which causes many privacy risks to be overlooked, especially data inferences. 

In addition to the participant's awareness and definition of privacy and security, the available rewards can influence how accepting the individual is towards collecting different data modalities. Our findings show that as the sensor becomes more intrusive and the data becomes more sensitive, the greater the incentive must be. In the interview, there was a distinction between rewards: monetary compensation, well-being insights, and improving building automation as potential benefits. While previous research has shown monetary rewards can motivate many different behaviors, previous works have shown that monetary contributions may not cause people to overlook their privacy concerns or incentivize contributions\cite{Lee2015CompensationBehaviour,sharif2022work,woolley2021incentives,Cappa2019MonetaryCrowdsourcing}. However, it was clear in participant responses that significantly higher compensations are needed as an incentive for more intrusive modalities (i.e.,  A/V). It is difficult to quantify the loss of privacy as described by Figure \ref{fig:sensitivity}, and there is a limit to how much money is available for compensation. In contrast, large societal impact and insights for improved well-being were influential motives for sharing more sensitive data to a greater degree than monetary compensation, which is supported by a previous study with sharing dashcam video data \cite{Kim2020Dashcam}. It is important to note that some responses from our study highlight that they would still like to know how a specific data modality will provide the insight and whether a less sensitive data modality could provide the same utility with an acceptable margin of loss of information.

Though many participants prefer to spend as little as possible on privacy settings, a few indicated that they were willing to spend from 30 minutes to one to two hours. However, quantifying time is difficult, and people may not want to spend as much time as they specified. Future time perception has been found to influence decision-making and behavior, especially when a future time interval ends in a gain or loss \cite{Bilgin2010LoomingPerception}. Depending on how an individual may perceive the task of protecting privacy as a chore ("negative") as compared to a beneficial one ("positive"), the time may be longer or shorter. People do care about privacy but do not want to spend time as it should be a given. There have been studies on preserving privacy in sensor data, primarily on handling data after collection and not prior \cite{Pappachan2017TowardsPreferences, Yang2021TowardsIoT, Koh2019WhoApplications}. Thus, a solution could be personalized default privacy settings that make protecting one's privacy as intuitive as possible and user-centric. This study shows that privacy preferences are context-specific, so we need flexible systems to adhere to individual preferences.

A limitation of our study was that the participants were highly educated with a specific background in cyber-physical systems, including smart cities and buildings. Since there were significant gaps between the perceptions and reality of smart building privacy found in this participant pool, it warrants further research into people who may not necessarily work in a smart building. Other studies \cite{Madden2017DigitalInequality, Zukowski2007DemographicsPrivacy} have found that individual differences, such as demographics and cultural differences, can contribute to different privacy concerns and preferences. The potential influence of demographics on people's privacy preferences in smart buildings warrants in-depth research as current literature broadly focuses on online privacy.

\begin{figure}[ht]
  \centering
  \includegraphics[width=0.70\linewidth]{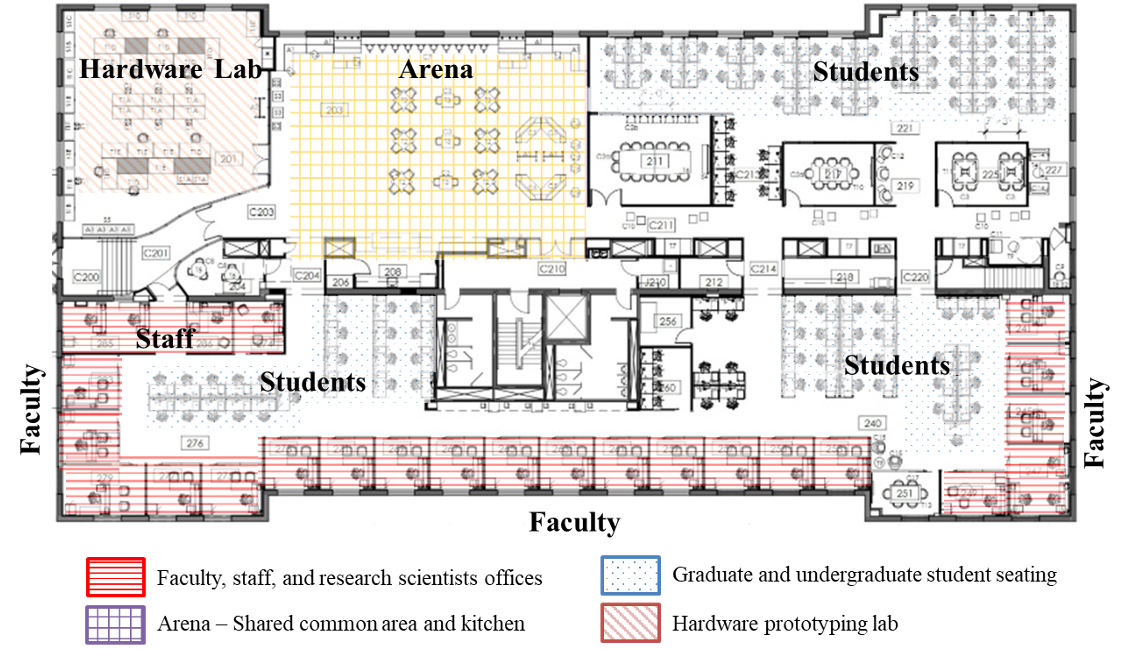}
  \caption{The overview of the smart office building space that participants work in.}
  \label{fig:llmap}
\end{figure}

\section*{Methods}\label{method}
We conducted semi-structured interviews with twenty-four residents of a smart office building situated within a university as described by Figure \ref{fig:llmap}. The interviews aim to understand the residents' awareness, perception, and data privacy preferences within a smart office building space. The study was approved by the Institutional Review Board for Social \& Behavioral Sciences at the University of Virginia (IRB-SBS Protocol \#4975). All participants reviewed the informed consent and agreed to participate in the study. All methods were performed in accordance with the relevant guidelines and regulations.

The targeted space comprises of more than 250 residents (faculty and graduate research assistants). The space is designed similarly to an open space commercial office building with individual cubicles, conference rooms, offices, and a common space. This space is a Cyber-Physical-Systems (CPS) research facility. As a result, the residents are familiar with the different areas of CPS, security, and policy, so we consider the group to be highly educated about autonomous systems, hardware for IoT, smart cities, and smart health.   

\subsection*{Participants}
A recruitment email was sent via the listserv for residents of the smart office space, which includes students and staff. Twenty-four participants were interviewed between April and May 2022. Most participants (83.3\%) were graduate students, and the remaining were staff and faculty at the University of Virginia, representing the overall population of the space. We reached data saturation after interviewing these twenty-four participants. There were 11 female and 11 male participants, while two participants preferred not to disclose their gender identity. The age distribution of the participants was as follows: 18 to 24 years (\textit{n}=4), 25-34 years (\textit{n}=17), 45-54 years (\textit{n}=1), and preferred not to answer (\textit{n}=1). 

\subsection*{Interview Procedure}\label{procedure}
They were asked to answer a few demographic questions such as gender, age, and role at the University. Participants could choose to have the interview online, via Zoom, or in person, where the audio was recorded. Each participant was compensated with a \$10 gift card for participating in this study. Interviews were semi-structured and focused on questions as listed:
\begin{enumerate}
    \item What smart devices do you own?
    \item What were your criteria in deciding what devices to buy?
    \item Could you point out the areas that you frequent?
    \item Which areas would you care about data being collected? Why?
    \item What data do you think is being collected in the [name of workplace]?
    \item Who do you think has access to the data? 
    \item Under what circumstances would you not be okay with data being collected and analyzed?
    \item What if you benefit from data being collected?
    \item In the context of smart buildings and the work environment, what is your definition of privacy? security? 
    \item Would you be willing to take steps to protect your privacy?
\end{enumerate}

The flow of the questions was designed not to prime the participants towards privacy-related answers and to mitigate biased answers, so the word privacy was not mentioned until the end of the interview. Two interviewers participated in each interview and worked together to follow up on points of interest that arose in conversation and ensure all interview questions were answered. Example points of interest were when participants mentioned privacy early on in the interviews or when the subject of identifiability of occupants was brought up. Participants were asked about IoT devices they own and their purchasing criteria to assess familiarity with IoTs. The rest of the interview focused on their experiences with IoTs within the smart office building at the University as well as their preferences, comfort, boundaries, and awareness pertaining to the collection of different data modalities -- audio, video, and environmental. We also asked for each individual's definition of privacy and security in the context of a smart building, such as the one they occupy for work, and their willingness to take steps to ensure their privacy.

\begin{figure}[ht]
  \centering
  \includegraphics[width=0.75\linewidth]{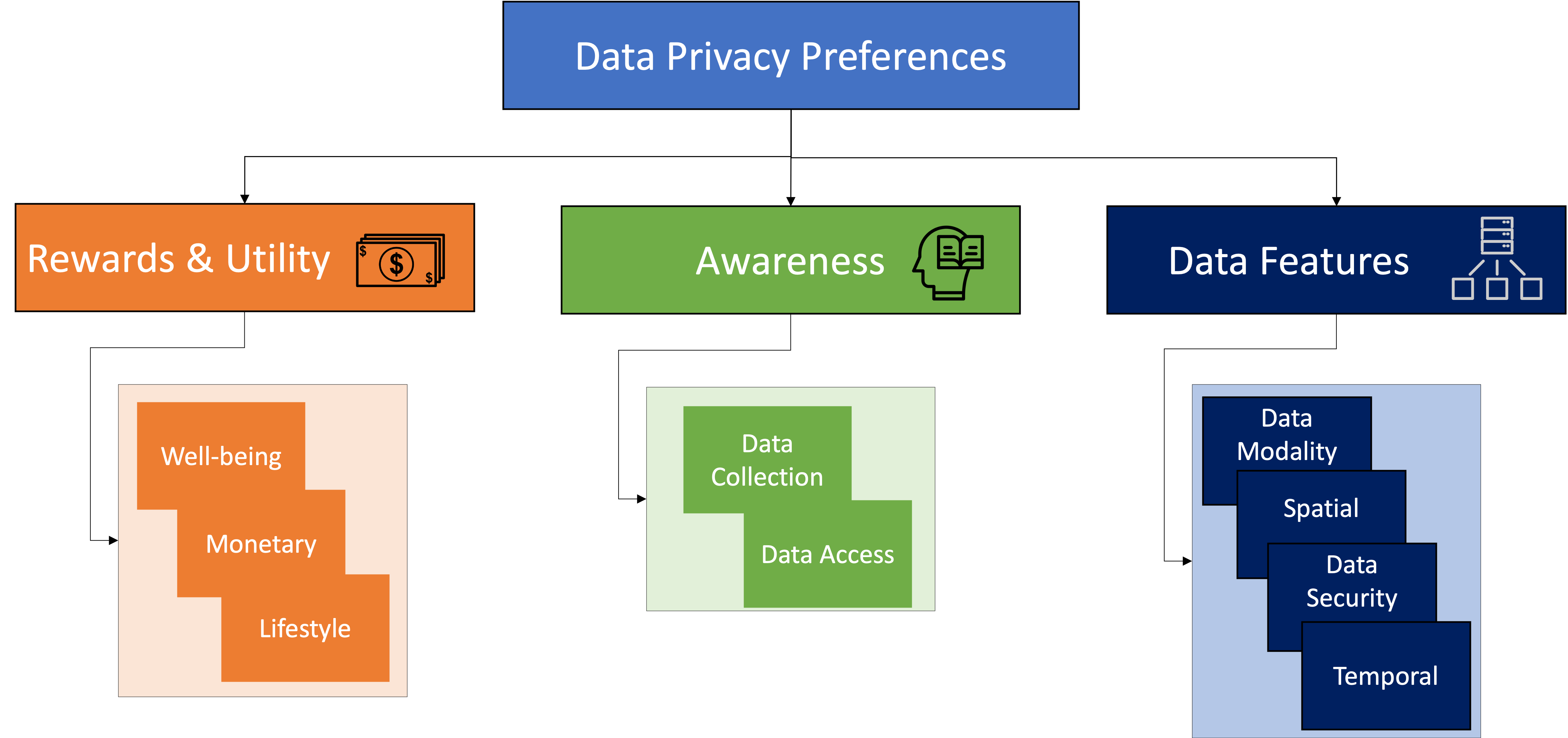}
  \caption{Interview responses were analyzed and categorized with codes. The initial codes were pooled from the separate coding of three researchers and refined through discussion and iterative analysis of the transcripts. The individual codes formed three categories -- \textit{Rewards \& Utility, User Knowledge, and data features} -- that fall under a larger umbrella of data privacy preferences. }
  \label{fig:codes}
\end{figure}

\subsection*{Analysis}\label{analysis}
We followed a standard procedure for analyzing qualitative data from interviews described by Seidman \cite{Seidman2019}. In summary, two researchers independently coded each transcription of the interviews, while a third researcher reviewed the codes and assisted in resolving conflicting opinions. A preliminary codebook is formed, which is a list of the words and phrases that encapsulate the data privacy preferences of users and help answer the research questions. The transcripts are then iteratively reviewed line by line and coded, then discussed among the research group to form the final code book. The final codebook consisted of three high-level categories, \textit{Rewards \& Utility, User Knowledge, and Data Modality Features}, with two to four codes each, as described by Figure \ref{fig:codes}. The final codebook was used to analyze and re-code the interviews to align with the refined codes. Through the three categories and the contained codes, several themes emerged. Any conflicts and questions from coding the interviews were resolved through discussions between three researchers.

\section*{Data Availability}
The datasets generated and/or analyzed during the current study are not publicly available following the IRB guidelines associated with this study but are available from the corresponding author on reasonable request.

\bibliography{references}

\begin{thebibliography}{10}
\urlstyle{rm}
\expandafter\ifx\csname url\endcsname\relax
  \def\url#1{\texttt{#1}}\fi
\expandafter\ifx\csname urlprefix\endcsname\relax\def\urlprefix{URL }\fi
\expandafter\ifx\csname doiprefix\endcsname\relax\def\doiprefix{DOI: }\fi
\providecommand{\bibinfo}[2]{#2}
\providecommand{\eprint}[2][]{\url{#2}}

\bibitem{Saputro2016PrivacyBuildings}
\bibinfo{author}{Saputro, N.}, \bibinfo{author}{Yurekli, A.~I.},
  \bibinfo{author}{Akkaya, K.} \& \bibinfo{author}{Uluagac, A.~S.}
\newblock \bibinfo{title}{{Privacy Preservation for IoT Used in Smart
  Buildings}}.
\newblock In \emph{\bibinfo{booktitle}{Security and Privacy in Internet of
  Things (IoTs): Models, Algorithms, and Implementations}},
  chap.~\bibinfo{chapter}{7}, \bibinfo{pages}{129--160},
  \doiprefix\url{10.1201/b19516-10} (\bibinfo{publisher}{CRC Press},
  \bibinfo{year}{2016}), \bibinfo{edition}{1} edn.

\bibitem{Labeodan2015OccupancyEvaluation}
\bibinfo{author}{Labeodan, T.}, \bibinfo{author}{Zeiler, W.},
  \bibinfo{author}{Boxem, G.} \& \bibinfo{author}{Zhao, Y.}
\newblock \bibinfo{journal}{\bibinfo{title}{{Occupancy measurement in
  commercial office buildings for demand-driven control applications—A survey
  and detection system evaluation}}}.
\newblock {\emph{\JournalTitle{Energy and Buildings}}}
  \textbf{\bibinfo{volume}{93}}, \bibinfo{pages}{303--314},
  \doiprefix\url{10.1016/J.ENBUILD.2015.02.028} (\bibinfo{year}{2015}).

\bibitem{Molina-Solana2017DataReview}
\bibinfo{author}{Molina-Solana, M.}, \bibinfo{author}{Ros, M.},
  \bibinfo{author}{Ruiz, M.~D.}, \bibinfo{author}{G{\'{o}}mez-Romero, J.} \&
  \bibinfo{author}{Martin-Bautista, M.~J.}
\newblock \bibinfo{journal}{\bibinfo{title}{{Data science for building energy
  management: A review}}}.
\newblock {\emph{\JournalTitle{Renewable and Sustainable Energy Reviews}}}
  \textbf{\bibinfo{volume}{70}}, \bibinfo{pages}{598--609},
  \doiprefix\url{10.1016/J.RSER.2016.11.132} (\bibinfo{year}{2017}).

\bibitem{Awada2021HomeIEQWorker}
\bibinfo{author}{Awada, M.}, \bibinfo{author}{Becerik-Gerber, B.},
  \bibinfo{author}{Lucas, G.} \& \bibinfo{author}{Roll, S.~C.}
\newblock \bibinfo{journal}{\bibinfo{title}{Associations among home indoor
  environmental quality factors and worker health while working from home
  during covid-19 pandemic}}.
\newblock {\emph{\JournalTitle{Journal of Engineering for Sustainable Buildings
  and Cities}}} \textbf{\bibinfo{volume}{2}}, \doiprefix\url{10.1115/1.4052822}
  (\bibinfo{year}{2021}).

\bibitem{Aryal2018SmartDesk}
\bibinfo{author}{Aryal, A.}, \bibinfo{author}{Anselmo, F.} \&
  \bibinfo{author}{Becerik-Gerber, B.}
\newblock \bibinfo{title}{Smart iot desk for personalizing indoor environmental
  conditions}.
\newblock \doiprefix\url{10.1145/3277593.3277614}
  (\bibinfo{publisher}{Association for Computing Machinery},
  \bibinfo{year}{2018}).

\bibitem{Aryal2018EnergyComfortOffice}
\bibinfo{author}{Aryal, A.} \& \bibinfo{author}{Becerik-Gerber, B.}
\newblock \bibinfo{journal}{\bibinfo{title}{Energy consequences of
  comfort-driven temperature setpoints in office buildings}}.
\newblock {\emph{\JournalTitle{Energy and Buildings}}}
  \textbf{\bibinfo{volume}{177}}, \bibinfo{pages}{33--46},
  \doiprefix\url{10.1016/J.ENBUILD.2018.08.013} (\bibinfo{year}{2018}).

\bibitem{Simin2018ActivityEnergyBuild}
\bibinfo{author}{Ahmadi-Karvigh, S.}, \bibinfo{author}{Ghahramani, A.},
  \bibinfo{author}{Becerik-Gerber, B.} \& \bibinfo{author}{Soibelman, L.}
\newblock \bibinfo{journal}{\bibinfo{title}{Real-time activity recognition for
  energy efficiency in buildings}}.
\newblock {\emph{\JournalTitle{Applied Energy}}}
  \textbf{\bibinfo{volume}{211}}, \bibinfo{pages}{146--160},
  \doiprefix\url{10.1016/J.APENERGY.2017.11.055} (\bibinfo{year}{2018}).

\bibitem{Ghahramani2018UnsupervisedThermal}
\bibinfo{author}{Ghahramani, A.}, \bibinfo{author}{Castro, G.},
  \bibinfo{author}{Karvigh, S.~A.} \& \bibinfo{author}{Becerik-Gerber, B.}
\newblock \bibinfo{journal}{\bibinfo{title}{Towards unsupervised learning of
  thermal comfort using infrared thermography}}.
\newblock {\emph{\JournalTitle{Applied Energy}}}
  \textbf{\bibinfo{volume}{211}}, \bibinfo{pages}{41--49},
  \doiprefix\url{10.1016/J.APENERGY.2017.11.021} (\bibinfo{year}{2018}).

\bibitem{Becerik-Gerber2022TenQuestionsHBI}
\bibinfo{author}{Becerik-Gerber, B.} \emph{et~al.}
\newblock \bibinfo{journal}{\bibinfo{title}{Ten questions concerning
  human-building interaction research for improving the quality of life}}.
\newblock {\emph{\JournalTitle{Building and Environment}}}
  \textbf{\bibinfo{volume}{226}},
  \doiprefix\url{10.1016/J.BUILDENV.2022.109681} (\bibinfo{year}{2022}).

\bibitem{Becerik-Gerber2022HBI}
\bibinfo{author}{Becerik-Gerber, B.} \emph{et~al.}
\newblock \bibinfo{journal}{\bibinfo{title}{The field of human building
  interaction for convergent research and innovation for intelligent built
  environments}}.
\newblock {\emph{\JournalTitle{Scientific Reports}}}
  \textbf{\bibinfo{volume}{12}}, \bibinfo{pages}{1--19},
  \doiprefix\url{10.1038/s41598-022-25047-y} (\bibinfo{year}{2022}).

\bibitem{Abade2018AEnvironments}
\bibinfo{author}{Abade, B.}, \bibinfo{author}{Abreu, D.~P.} \&
  \bibinfo{author}{Curado, M.}
\newblock \bibinfo{journal}{\bibinfo{title}{{A Non-Intrusive Approach for
  Indoor Occupancy Detection in Smart Environments}}}.
\newblock {\emph{\JournalTitle{Sensors (Basel, Switzerland)}}}
  \textbf{\bibinfo{volume}{18}}, \doiprefix\url{10.3390/S18113953}
  (\bibinfo{year}{2018}).

\bibitem{Weng2012FromEfficiency}
\bibinfo{author}{Weng, T.} \& \bibinfo{author}{Agarwal, Y.}
\newblock \bibinfo{journal}{\bibinfo{title}{{From Buildings to Smart
  Buildings—Sensing and Actuation to Improve Energy Efficiency}}}.
\newblock {\emph{\JournalTitle{IEEE Design and Test of Computers}}}
  \textbf{\bibinfo{volume}{29}}, \bibinfo{pages}{36--44},
  \doiprefix\url{10.1109/MDT.2012.2211855} (\bibinfo{year}{2012}).

\bibitem{Lee2019ExploringBuildings}
\bibinfo{author}{Lee, P.} \emph{et~al.}
\newblock \bibinfo{title}{{Exploring Privacy Breaches and Mitigation Strategies
  of Occupancy Sensors in Smart Buildings}}.
\newblock In \emph{\bibinfo{booktitle}{TESCA'19: Proceedings of the 1st ACM
  International Workshop on Technology Enablers and Innovative Applications for
  Smart Cities and Communities}}, \bibinfo{pages}{18--21},
  \doiprefix\url{10.1145/3364544} (\bibinfo{publisher}{Association for
  Computing Machinery}, \bibinfo{address}{New York, NY, USA},
  \bibinfo{year}{2019}).

\bibitem{Kroger2019UnexpectedThings}
\bibinfo{author}{Kr{\"{o}}ger, J.}
\newblock \bibinfo{title}{{Unexpected Inferences from Sensor Data: A Hidden
  Privacy Threat in the Internet of Things}}.
\newblock In \bibinfo{editor}{Strous, L.} \& \bibinfo{editor}{Cerf, V.~G.}
  (eds.) \emph{\bibinfo{booktitle}{IFIP Advances in Information and
  Communication Technology}}, vol. \bibinfo{volume}{548},
  \bibinfo{pages}{147--159}, \doiprefix\url{10.1007/978-3-030-15651-0{\_}13}
  (\bibinfo{publisher}{Springer, Cham}, \bibinfo{year}{2019}).

\bibitem{Harper2020UserBuildings}
\bibinfo{author}{Harper, S.}, \bibinfo{author}{Mehrnezhad, M.} \&
  \bibinfo{author}{Mace, J.~C.}
\newblock \bibinfo{title}{{User Privacy Concerns and Preferences in Smart
  Buildings}}.
\newblock In \emph{\bibinfo{booktitle}{Socio-Technical Aspects in Security and
  Trust}}, vol. \bibinfo{volume}{12812 LNCS}, \bibinfo{pages}{85--106},
  \doiprefix\url{10.1007/978-3-030-79318-0{\_}5} (\bibinfo{publisher}{Springer
  International Publishing}, \bibinfo{year}{2020}).

\bibitem{Apthorpe2018DiscoveringIntegrity}
\bibinfo{author}{Apthorpe, N.}, \bibinfo{author}{Shvartzshnaider, Y.},
  \bibinfo{author}{Mathur, A.}, \bibinfo{author}{Reisman, D.} \&
  \bibinfo{author}{Feamster, N.}
\newblock \bibinfo{journal}{\bibinfo{title}{{Discovering Smart Home Internet of
  Things Privacy Norms Using Contextual Integrity}}}.
\newblock {\emph{\JournalTitle{Proceedings of the ACM on Interactive, Mobile,
  Wearable and Ubiquitous Technologies}}} \textbf{\bibinfo{volume}{2}},
  \bibinfo{pages}{1--23}, \doiprefix\url{10.1145/3214262}
  (\bibinfo{year}{2018}).

\bibitem{Pathmabandu2020AnBuildings}
\bibinfo{author}{Pathmabandu, C.}, \bibinfo{author}{Grundy, J.},
  \bibinfo{author}{Chhetri, M.~B.} \& \bibinfo{author}{Baig, Z.}
\newblock \bibinfo{title}{{An Informed Consent Model for Managing the Privacy
  Paradox in Smart Buildings}}.
\newblock In \emph{\bibinfo{booktitle}{Proceedings of the 35th IEEE/ACM
  International Conference on Automated Software Engineering Workshops}},
  \bibinfo{pages}{19--26}, \doiprefix\url{10.1145/3417113.3422180}
  (\bibinfo{publisher}{Association for Computing Machinery},
  \bibinfo{address}{New York, NY, USA}, \bibinfo{year}{2020}).

\bibitem{Wu2021TheChallenge}
\bibinfo{author}{Wu, T.} \emph{et~al.}
\newblock \bibinfo{title}{{The Smart Building Privacy Challenge}}.
\newblock In \emph{\bibinfo{booktitle}{BuildSys '21: Proceedings of the 8th ACM
  International Conference on Systems for Energy-Efficient Buildings, Cities,
  and Transportation}}, \bibinfo{pages}{238--239},
  \doiprefix\url{10.1145/3486611.3492234} (\bibinfo{publisher}{Association for
  Computing Machinery}, \bibinfo{address}{Coimbra, Portugal},
  \bibinfo{year}{2021}).

\bibitem{Gorjani2022SmartHomeActivity}
\bibinfo{author}{Gorjani, O.~M.}, \bibinfo{author}{Bilik, P.} \&
  \bibinfo{author}{Koziorek, J.}
\newblock \bibinfo{journal}{\bibinfo{title}{Activity recognition within smart
  homes using logistic regression}}.
\newblock {\emph{\JournalTitle{14th International Conference ELEKTRO, ELEKTRO
  2022 - Proceedings}}} \doiprefix\url{10.1109/ELEKTRO53996.2022.9803583}
  (\bibinfo{year}{2022}).

\bibitem{Kessler2020PrivacyBuildings}
\bibinfo{author}{Kessler, E.}, \bibinfo{author}{Masiane, M.} \&
  \bibinfo{author}{Abdelhalim, A.}
\newblock \bibinfo{journal}{\bibinfo{title}{Privacy concerns regarding occupant
  tracking in smart buildings}}.
\newblock {\emph{\JournalTitle{arXiv preprint arXiv:2010.07028}}}
  (\bibinfo{year}{2020}).

\bibitem{Zheng2018UserPrivacy}
\bibinfo{author}{Zheng, S.}, \bibinfo{author}{Apthorpe, N.},
  \bibinfo{author}{Chetty, M.} \& \bibinfo{author}{Feamster, N.}
\newblock \bibinfo{journal}{\bibinfo{title}{{User Perceptions of Smart Home IoT
  Privacy}}}.
\newblock {\emph{\JournalTitle{Proceedings of the ACM on Human-Computer
  Interaction}}} \textbf{\bibinfo{volume}{2}}, \bibinfo{pages}{1--20},
  \doiprefix\url{10.1145/3274469} (\bibinfo{year}{2018}).

\bibitem{McCreary2016TheBuildings}
\bibinfo{author}{McCreary, F.}, \bibinfo{author}{Zafiroglu, A.} \&
  \bibinfo{author}{Patterson, H.}
\newblock \bibinfo{title}{{The Contextual Complexity of Privacy in Smart Homes
  and Smart Buildings}}.
\newblock In \bibinfo{editor}{Nah, F. F.-H.} \& \bibinfo{editor}{Tan, C.-H.}
  (eds.) \emph{\bibinfo{booktitle}{HCI in Business, Government, and
  Organizations: Information Systems}}, vol. \bibinfo{volume}{9752},
  \bibinfo{pages}{67--78},
  \doiprefix\url{10.1007/978-3-319-39399-5{\_}7/FIGURES/4}
  (\bibinfo{publisher}{Springer, Cham}, \bibinfo{year}{2016}).

\bibitem{Harper2022CommercialSB}
\bibinfo{author}{Harper, S.}, \bibinfo{author}{Mehrnezhad, M.} \&
  \bibinfo{author}{Mace, J.}
\newblock \bibinfo{journal}{\bibinfo{title}{User privacy concerns in commercial
  smart buildings}}.
\newblock {\emph{\JournalTitle{Journal of Computer Security}}}
  \textbf{\bibinfo{volume}{30}}, \bibinfo{pages}{465--497},
  \doiprefix\url{10.3233/JCS-210035} (\bibinfo{year}{2022}).

\bibitem{Lee2018SensingUtility}
\bibinfo{author}{Lee, A.~J.}, \bibinfo{author}{Biehl, J.~T.} \&
  \bibinfo{author}{Curry, C.}
\newblock \bibinfo{title}{Sensing or watching? balancing utility and privacy in
  sensing systems via collection and enforcement mechanisms}.
\newblock \bibinfo{pages}{105--116}, \doiprefix\url{10.1145/3205977.3205983}
  (\bibinfo{publisher}{Association for Computing Machinery},
  \bibinfo{year}{2018}).

\bibitem{Nissenbaum2004LawReview}
\bibinfo{author}{Nissenbaum, H.}
\newblock \bibinfo{journal}{\bibinfo{title}{Privacy as contextual integrity}}.
\newblock {\emph{\JournalTitle{Washington Law Review}}}
  \textbf{\bibinfo{volume}{79}}, \bibinfo{pages}{2--3} (\bibinfo{year}{2004}).

\bibitem{Nissenbaum2009PrivacyLife}
\bibinfo{author}{Nissenbaum, H.}
\newblock \emph{\bibinfo{title}{{Privacy in Context: Technology, Policy, and
  the Integrity of Social Life}}} (\bibinfo{publisher}{Stanford Law Books},
  \bibinfo{year}{2009}).

\bibitem{Marcello2019Recognition}
\bibinfo{author}{Marcello, F.} \& \bibinfo{author}{Pilloni, V.}
\newblock \bibinfo{journal}{\bibinfo{title}{Sensor-based activity recognition
  inside smart building energy and comfort management systems}}.
\newblock {\emph{\JournalTitle{IEEE 5th World Forum on Internet of Things,
  WF-IoT 2019}}} \bibinfo{pages}{639--643},
  \doiprefix\url{10.1109/WF-IOT.2019.8767233} (\bibinfo{year}{2019}).

\bibitem{Lee2015CompensationBehaviour}
\bibinfo{author}{Lee, H.}, \bibinfo{author}{Lim, D.}, \bibinfo{author}{Kim,
  H.}, \bibinfo{author}{Zo, H.} \& \bibinfo{author}{Ciganek, A.~P.}
\newblock \bibinfo{journal}{\bibinfo{title}{{Compensation paradox: the
  influence of monetary rewards on user behaviour}}}.
\newblock {\emph{\JournalTitle{Behaviour {\&} Information Technology}}}
  \textbf{\bibinfo{volume}{34}}, \bibinfo{pages}{45--56},
  \doiprefix\url{10.1080/0144929X.2013.805244} (\bibinfo{year}{2015}).

\bibitem{sharif2022work}
\bibinfo{author}{Sharif, M.~A.} \& \bibinfo{author}{Woolley, K.}
\newblock \bibinfo{journal}{\bibinfo{title}{Work-to-unlock rewards: Leveraging
  goals in reward systems to increase consumer persistence}}.
\newblock {\emph{\JournalTitle{Journal of Consumer Research}}}
  \textbf{\bibinfo{volume}{49}}, \bibinfo{pages}{634--656}
  (\bibinfo{year}{2022}).

\bibitem{woolley2021incentives}
\bibinfo{author}{Woolley, K.} \& \bibinfo{author}{Sharif, M.~A.}
\newblock \bibinfo{journal}{\bibinfo{title}{Incentives increase relative
  positivity of review content and enjoyment of review writing}}.
\newblock {\emph{\JournalTitle{Journal of Marketing Research}}}
  \textbf{\bibinfo{volume}{58}}, \bibinfo{pages}{539--558}
  (\bibinfo{year}{2021}).

\bibitem{Cappa2019MonetaryCrowdsourcing}
\bibinfo{author}{Cappa, F.}, \bibinfo{author}{Rosso, F.} \&
  \bibinfo{author}{Hayes, D.}
\newblock \bibinfo{journal}{\bibinfo{title}{{Monetary and Social Rewards for
  Crowdsourcing}}}.
\newblock {\emph{\JournalTitle{Sustainability}}} \textbf{\bibinfo{volume}{11}},
  \doiprefix\url{10.3390/SU11102834} (\bibinfo{year}{2019}).

\bibitem{Kim2020Dashcam}
\bibinfo{author}{Kim, J.}, \bibinfo{author}{Park, S.} \& \bibinfo{author}{Lee,
  U.}
\newblock \bibinfo{journal}{\bibinfo{title}{Dashcam witness: Video sharing
  motives and privacy concerns across different nations}}.
\newblock {\emph{\JournalTitle{IEEE Access}}} \textbf{\bibinfo{volume}{8}},
  \bibinfo{pages}{110425--110437}, \doiprefix\url{10.1109/ACCESS.2020.3002079}
  (\bibinfo{year}{2020}).

\bibitem{Bilgin2010LoomingPerception}
\bibinfo{author}{Bilgin, B.} \& \bibinfo{author}{LeBoeuf, R.~A.}
\newblock \bibinfo{journal}{\bibinfo{title}{{Looming Losses in Future Time
  Perception}}}.
\newblock {\emph{\JournalTitle{Journal of Marketing Research}}}
  \textbf{\bibinfo{volume}{47}}, \bibinfo{pages}{520--530},
  \doiprefix\url{10.1509/JMKR.47.3.520} (\bibinfo{year}{2010}).

\bibitem{Pappachan2017TowardsPreferences}
\bibinfo{author}{Pappachan, P.} \emph{et~al.}
\newblock \bibinfo{title}{{Towards Privacy-Aware Smart Buildings: Capturing,
  Communicating, and Enforcing Privacy Policies and Preferences}}.
\newblock In \emph{\bibinfo{booktitle}{2017 IEEE 37th International Conference
  on Distributed Computing Systems Workshops (ICDCSW)}},
  \bibinfo{pages}{193--198}, \doiprefix\url{10.1109/ICDCSW.2017.52}
  (\bibinfo{publisher}{Institute of Electrical and Electronics Engineers Inc.},
  \bibinfo{address}{Atlanta, GA, USA}, \bibinfo{year}{2017}).

\bibitem{Yang2021TowardsIoT}
\bibinfo{author}{Yang, X.}
\newblock \bibinfo{title}{{Towards utility-aware privacy-preserving sensor data
  anonymization in distributed IoT}}.
\newblock In \emph{\bibinfo{booktitle}{BuildSys '21: Proceedings of the 8th ACM
  International Conference on Systems for Energy-Efficient Buildings, Cities,
  and Transportation}}, \bibinfo{pages}{248--249},
  \doiprefix\url{10.1145/3486611.3492389} (\bibinfo{publisher}{Association for
  Computing Machinery}, \bibinfo{address}{Coimbra, Portugal},
  \bibinfo{year}{2021}).

\bibitem{Koh2019WhoApplications}
\bibinfo{author}{Koh, J.} \emph{et~al.}
\newblock \bibinfo{title}{{Who can Access What, and When? Understanding Minimal
  Access Requirements of Building Applications}}.
\newblock In \emph{\bibinfo{booktitle}{BuildSys '19: Proceedings of the 6th ACM
  International Conference on Systems for Energy-Efficient Buildings, Cities,
  and Transportation}}, \bibinfo{pages}{121--124},
  \doiprefix\url{10.1145/3360322.3360868} (\bibinfo{publisher}{Association for
  Computing Machinery}, \bibinfo{address}{New York, NY, USA},
  \bibinfo{year}{2019}).

\bibitem{Madden2017DigitalInequality}
\bibinfo{author}{Madden, M.}
\newblock \bibinfo{title}{Privacy, security, and digital inequality}
  (\bibinfo{year}{2017}).

\bibitem{Zukowski2007DemographicsPrivacy}
\bibinfo{author}{Zukowski, T.} \& \bibinfo{author}{Brown, I.}
\newblock \bibinfo{title}{Examining the influence of demographic factors on
  internet users' information privacy concerns}.
\newblock vol. \bibinfo{volume}{226}, \bibinfo{pages}{197--204},
  \doiprefix\url{10.1145/1292491.1292514} (\bibinfo{publisher}{Association for
  Computing Machinery}, \bibinfo{year}{2007}).

\bibitem{Seidman2019}
\bibinfo{author}{Seidman, I.}
\newblock \emph{\bibinfo{title}{Interviewing as Qualitative Research : A Guide
  for Researchers in Education and the Social Sciences}}
  (\bibinfo{year}{2019}), \bibinfo{edition}{5} edn.

\end{thebibliography}

\section*{Acknowledgements}
This paper is based upon work supported by the National Science Foundation (NSF) Research Traineeship (NRT) program under Grant No.182900 and NSF Grant No. 1823325.

\section*{Author contributions statement}
B.L. conceived and designed this study. B.L. and A.T. conducted the interviews. B.L., A.T., and A.H. reviewed and coded the interview transcripts. B.L. conducted the analysis and drafted the manuscript with revisions by A.T. and A.H. All authors reviewed the final manuscript.

\section*{Competing Interests}
The authors declare no competing interests.

\section*{Additional Information}
Correspondence and requests for materials should be addressed to A.H.


\end{document}